\documentclass[prl, twocolumn, amsmath, nofootinbib, floatfix, showpacs, preprintnumbers,letter]{revtex4} 
\usepackage{graphicx}
\input epsf
\usepackage{color,aas_macros}
\newcommand{\hMpc}{\,h^{-1}{\rm Mpc}}
\newcommand{\hMpcI}{\,h\,{\rm Mpc}^{-1}}
\newcommand{\vn}{\hat{n}}
\newcommand{\flux}{\mathcal{F}}
\def\lcdm{$\Lambda$CDM~}

\def\be{\begin{equation}}
\def\ee{\end{equation}}
\def\bea{\begin{eqnarray}}
\def\eea{\end{eqnarray}}
\newcommand{\vs}{\nonumber\\}
\newcommand{\Lya}{Lyman-$\alpha${ }}

\begin{document}

\preprint{FERMILAB-PUB-09-084-A}

\title{Lenses in the Forest: Cross--Correlation of the \Lya Flux\\with Cosmic Microwave Background Lensing}

 \author{Alberto Vallinotto$^{1,2}$}
 \author{Sudeep Das$^{3,4}$}
 \author{David N. Spergel$^{3,5}$}
 \author{Matteo Viel$^{6,7}$}

 \affiliation{$^1$Institut d'Astrophysique de Paris,
 CNRS-UMR 7095, Universit\'{e} Paris VI Pierre et Marie Curie,
 98 bis boulevard Arago, 75014 Paris, France.}
\affiliation{$^2$Center for Particle Astrophysics, Fermi National Accelerator Laboratory, 
 P.O. Box 500, Kirk Rd. \& Pine St., Batavia, IL 60510-0500 USA} 
 \affiliation{$^3$Princeton University Observatory,
Peyton Hall, Ivy Lane, Princeton, NJ 08544 USA}
\affiliation{$^4$Department of Physics, Jadwin Hall, Princeton University, Princeton, NJ 08544 USA}
 \affiliation{$^5$Astroparticule et Cosmologie APC, 10, rue Alice Domon et 
L\'eonie Duquet, 75205 Paris cedex 13, France}
 \affiliation{$^6$INAF - Osservatorio Astronomico di Trieste, Via G.B. Tiepolo 11,
I-34131 Trieste, Italy}
\affiliation{$^7$INFN - National Institute for Nuclear Physics, Via Valerio 2,
I-34127 Trieste, Italy}

\email{avalli@fnal.gov}
%\email{sudeep@astro.princeton.edu}
%\email{dns@astro.princeton.edu}
%\email{viel@oats.inaf.it}

\date{\today}
%\smallskip
\begin{abstract}
 We present a theoretical estimate for a new observable: the
 cross--correlation between the \Lya flux fluctuations
 in quasar (QSO) spectra and the convergence of the cosmic microwave
 background (CMB) as measured along the same line--of--sight. As a first
 step toward the assessment of its detectability, we estimate the
 signal--to--noise ratio using linear theory. Although the
 signal--to--noise is small for a single line--of--sight and peaks at
 somewhat smaller redshifts than those probed by the \Lya forest, we estimate  a total signal--to--noise of 9 for cross--correlating QSO spectra of SDSS-III with Planck and 20 for cross--correlating with a future polarization based CMB experiment. The detection of this effect would be a direct measure of the neutral hydrogen--matter cross--correlation and could provide important information on the growth of structures at
 large scales in a redshift range which is still poorly probed by
 observations.
\end{abstract}


\maketitle

\textit{Introduction.}  The \Lya forest -- the absorption seen in
quasar (QSO) spectra caused by intervening neutral hydrogen of the
intergalactic medium (IGM) -- has the potential to provide precise
information about the matter distribution down to small scales. In the
near future, the SDSS III Baryon Oscillation Spectroscopic Survey
(BOSS) will measure absorption spectra toward $160,000$ QSOs, a
fiftyfold improvement on existing surveys \cite{mcdonald05}.
This large sample could potentially provide unprecedented constraints
on neutrino masses \cite{gratton.lewis.ea:2008}, amplitude and slope
of the matter power spectrum \cite{lesgourgues07}, inflationary
parameters and the running spectral index
\cite{seljak.slosar:2006}. Furthermore, the baryon oscillations in the
\Lya forest can be used to constrain dark energy and curvature
\cite{mcdonald.eisentein:2007} and high-resolution QSO spectra can
also play a significant role in constraining the properties of
dark matter on small scales \cite{viel08}.  

The theoretical studies
assume that the physics of the \Lya forest is relatively simple and
that the observed flux fluctuations faithfully trace the dark matter
distribution.  Hydrodynamical simulations play an essential role in
testing this assumption through comparisons with data and in
calibrating the measurements. However, many systematic errors still
pose challenges at the sub-percent level of accuracy required by the
data. The main sources of uncertainty include, but are not limited to:
the accurate modeling of fluctuations in the ionization background,
the QSO continuum fluctuations, the uncertainties in the slope of the
temperature-density relation in the IGM and fluctuations about this mean relation, the effect of galactic
superwinds on the \Lya flux and the contamination of the \Lya spectrum
with metal lines (e.g.~\cite{vhs04}). The
ingredients that go into the simulations could be better controlled if
we could directly and independently test how the fluctuations in the
transmitted flux and the dark matter are related. 

In this \textit{Letter}, we
propose one such observable: the cross--correlation of the \Lya flux
fluctuations with the convergence field extracted from the
gravitational lensing of the cosmic microwave background (CMB).
Gravitational lensing of the CMB is caused by the deflection of CMB
photons by intervening large scale structure \cite{lewis.challinor:2006}.  Lensing breaks the statistical
isotropy of the CMB, introduces non-Gaussianities and produces B-mode
polarization from E-modes. Observations of small scale temperature
and polarization fluctuations can be used to reconstruct the effective
deflection field that lensed the CMB
\cite{hu.okamoto:2002}. WMAP
measurements have enabled the first detection of CMB
lensing \cite{smith.zahn.ea:2007, hirata08}. With the ongoing
and upcoming high resolution temperature and polarization based CMB
experiments like the Atacama Cosmology Telescope, the South
Pole Telescope, Planck, QUIET 
and PolarBearR \cite{ALL}, high fidelity lensing reconstruction will
soon become a reality. The timely convergence of the \Lya surveys and
high resolution CMB experiments in coming few years will make the
measurement of their cross--correlation a possibility. 

We note that
cross--correlating CMB temperature maps with the large-scale structure
(LSS) is currently done with many LSS probes including \Lya
(e.g. \cite{peiris.spergel:2000}), however we
focus here on the \textit{convergence} of the CMB rather than the CMB
temperature fluctuations. The cross--correlation signal between these
two quantities should provide insights on the relative bias
between matter and flux (in a way somewhat analogous to
\cite{hirata08}) and may 
become a powerful tool to calibrate the relationship
between the IGM and the dark matter. As a first step toward
understanding the information content and the detectability of this
correlation, we provide here a simple theoretical estimate of the
signal and of the significance with which it can be measured by
the upcoming experiments.

%In this work, we aim at cross-correlating at the largest scales the
%one--dimensional flux distribution of \Lya QSO spectra with the
%convergence field extracted from the CMB: this should be regarded as a
%new preliminary attempt to combine these two different observables
%within the framework of some standard or alternative cosmological
%models. This work, although presented in a general way, is tailored to
%be applied on future data sets when both the number of QSOs will be
%larger and higher resolution CMB convergence maps will be available.

                                % and
                                % to %cross check and improve
                                % numerical simulations

\begin{figure*}[t!]
\includegraphics[width=0.45\textwidth]{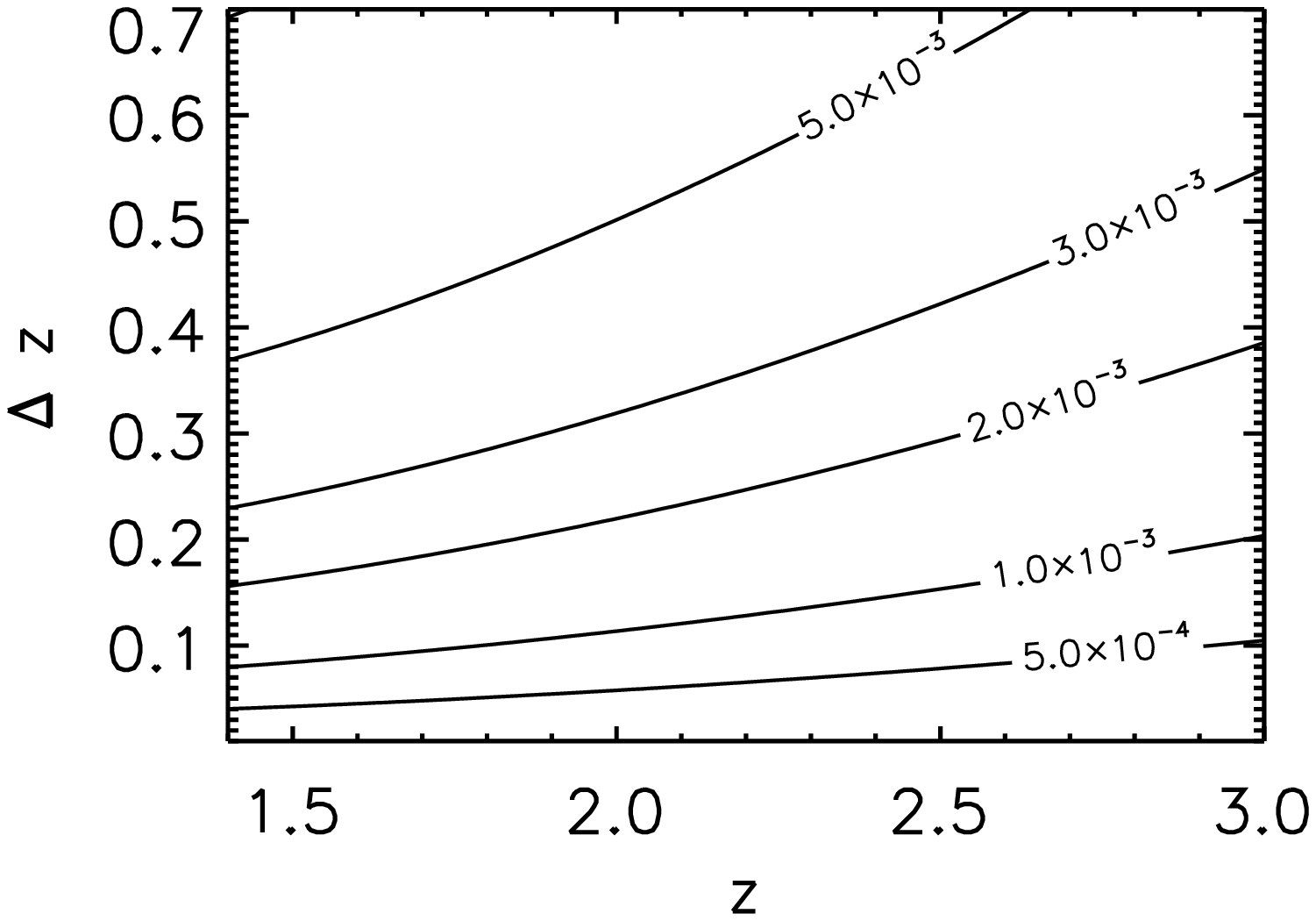}
\includegraphics[width=0.45\textwidth]{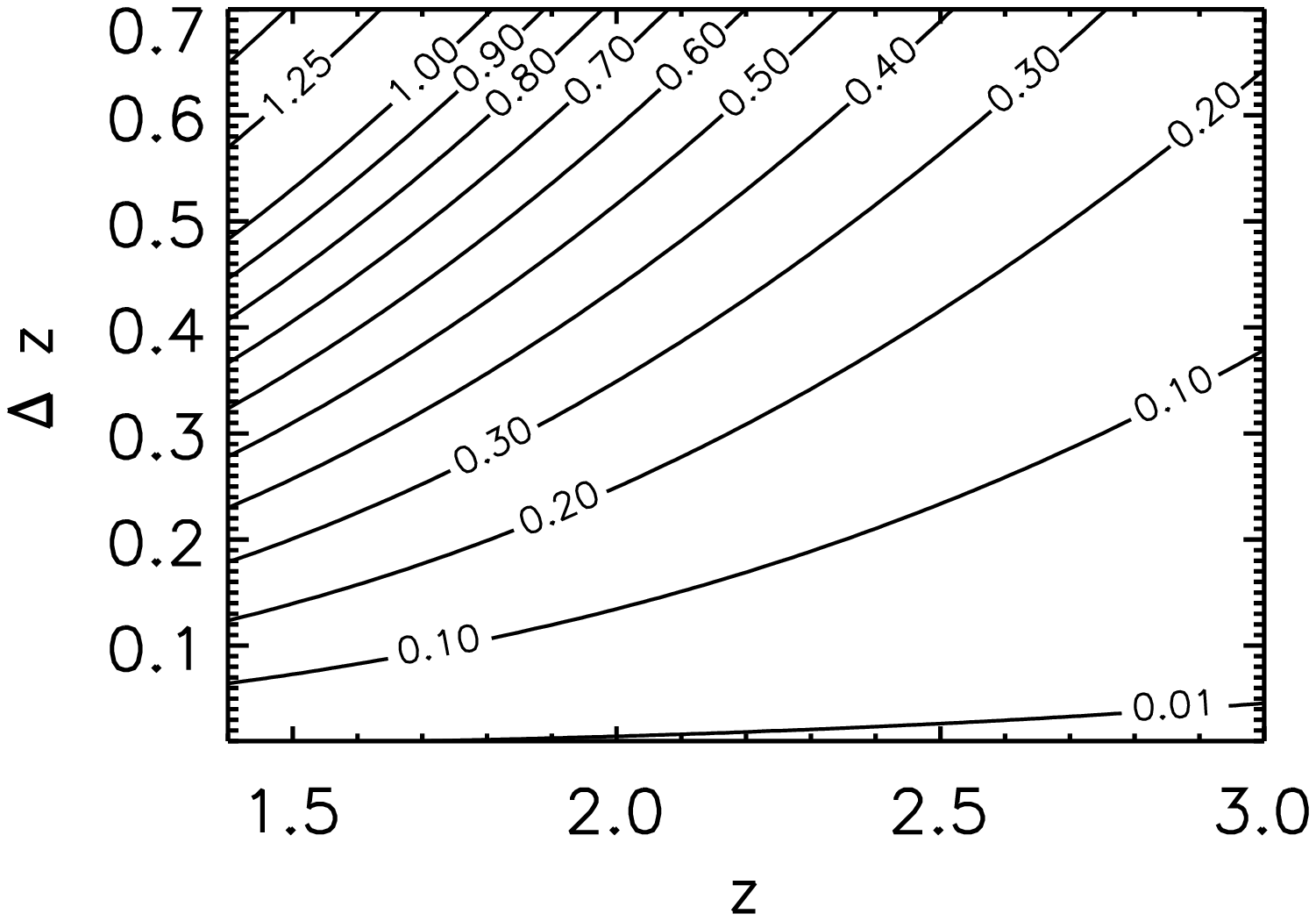}
\caption{Values of the correlators $\langle \delta\flux \kappa\rangle$
  (left panel) and $\langle \delta \flux^2 \kappa\rangle$ (right
  panel) along a \textit{single} \textit{los} as a function of the source
  redshift $z$ and of the length of the measured spectrum $\Delta
  z$. The cutoff scales assumed are $k_l=0.01\hMpcI$ and
  $k_L=4.8\hMpcI$ for the QSO spectrum and $k_C=0.021\hMpcI$ for the
  CMB convergence, consistent with SDSS-III and Planck
  specifications. We set $A=\beta=1$ so that it is straightforward to rescale the present figures to match the measured values of such parameters.}
\label{Fig:S_3_2}
\end{figure*}

\textit{Evaluation of the correlators.} We calculate the correlations
between the fluctuations in the flux $\delta\flux(\vn,\chi)$ and their
variance $\delta\flux^2(\vn,\chi)$ measured for a single QSO -- whose
line--of--sight (\textit{los}) lies in the direction $\vn$ and whose spectrum
covers a redshift range $\Delta z$ -- and the CMB convergence
$\kappa(\vn)$ measured in the \textit{same} direction. While the
convergence measures the projected matter fluctuations along the \textit{los},
the quantity $\delta\flux$ is sensitive to the typical fluctuations of
the \Lya forest which are in the range 1-100 comoving $\hMpc$. 
%Because the presence of dark matter will impact more directly on the variance of the fluctuations, the cross--correlation of the convergence with the flux \textit{variance} is expected to increase the signal. 
Since structures grow at a faster rate in overdense regions (which in turn are characterized by a positive convergence) \cite{Zaldarriaga:2000}, the cross--correlation of the convergence with the flux \textit{variance} is expected to increase the signal.
Also, using the flux variance makes the observable less sensitive to the continuum fitting uncertainties of the QSO spectrum since $\delta \flux$ requires correct extrapolation of the continuum from longward of \Lya.

%Using
%the variance of the flux fluctuation instead of the fluctuation itself
%in this cross--correlation is a way of both increasing the signal --
%since the presence of dark matter will impact more directly on the
%fluctuations variance -- and of being less sensitive to the continuum
%fitting uncertainties of the QSO spectrum.

The effective CMB convergence measured in the
direction $\vn$ depends on the amplitude of matter density fluctuations $\delta(\vn,\chi)$ along the los and reads:
\begin{equation}
\kappa(\hat{n})=\frac{3H_0^2\Omega_{\rm 0m}}{2c^2}\int_0^{\chi_{LS}}d\chi\,W_L(\chi,\chi_{LS})
\frac{\delta(\hat{n},\chi)}{a(\chi)}\label{kappa},
\end{equation}
where the integral is along the \textit{los}, $\chi_{LS}$ is the comoving distance to the last scattering surface,
 $W_L(\chi,\chi_{LS})=\chi(\chi_{LS}-\chi)/\chi_{LS}$ is the
usual lensing window function and $a(\chi)$ is the scale factor at comoving distance $\chi$.

We assume that on large scales fluctuations in the optical depth are
linearly proportional to fluctuations in the matter density
\cite{huignedin97}, motivated by the so--called fluctuating
Gunn--Peterson approximation \cite{Gunn:1965hd, Croft:1997jf} that relates flux and matter
$F=\exp{[-A(1+\delta)^\beta}]$, where $A$ and $\beta$ are two redshift
dependent functions: $A$ is of order unity and is related to the mean
flux level, baryon fraction, IGM temperature, cosmological parameters
and the photoionization rate of hydrogen, while $\beta$ depends on to
the so-called IGM temperature-density relation
(e.g.~\cite{huignedin97, mcdonald03}). However, for the purpose of this \textit{Letter}, it is fair to neglect their evolution with redshift and
to assume that they are constant over the observed QSO redshift
interval in each \textit{los}. In the large scale limit, the exponent is small
and we can expand it. Keeping only up to the lowest order term in
$\delta$ \cite{Viel:2001hd, Croft:1997jf} we get:
%\begin{equation}
 $\delta\flux (\vn,\chi) \approx- A \beta\delta(\vn,\chi)$. %+ \frac{A^2\beta^2}{2} \delta^2(\vn,\chi). 
% \label{deltaFExp}
%\end{equation} 
We stress that this approximation is valid in linear theory
neglecting not only the non--linearities produced by gravitational
instabilities but also those introduced by the definition of the flux,
thermal broadening and peculiar velocities. Higher order terms in the
expansion will, in general, contribute to the
cross--correlations being estimated here. To keep the calculations
analytically tractable, in this \textit{Letter} we neglect these higher order terms.

We define the $m-$th moment of the flux fluctuation averaged over the \textit{los} interval probed by the \Lya spectrum (extending from $\chi_i$ to $\chi_Q$) as:
\begin{eqnarray}
 \delta\flux^m(\vn)&=&\int_{\chi_i}^{\chi_Q}d\chi\, \delta\flux^m(\vn,\chi)
\vs&\approx&
 (-A\beta)^m\int_{\chi_i}^{\chi_Q}d\chi\, \delta^m(\vn,\chi)
\label{deltaF2}.
\end{eqnarray}
The correlation of this quantity with the CMB convergence $\kappa$
measured along the same \textit{los} is then,
 \begin{eqnarray}
 \langle \delta\flux^m(\hat{n}) \kappa(\hat{n})\rangle&=&(-A\beta)^m\frac{3H_0^2\Omega_m}{2c^2}
\int_0^{\chi_{LS}}d\chi_c 
\frac{W_L(\chi_c,\chi_{LS})}{a(\chi_c)}
\vs&\times& 
\int_{\chi_i}^{\chi_Q}d\chi_q \,
\langle \delta^m(\hat{n},\chi_q) \, \delta(\hat{n},\chi_c) \rangle.\label{deltaF2K_1}
\end{eqnarray}
We focus here on the case $m=1$ (and $m=2$), which is the
cross--correlation between the CMB convergence and the QSO mean flux fluctuation (and variance). To proceed further we Fourier transform the cumulant
correlator $\langle \delta^m(\hat{n},\chi_q) \, \delta(\hat{n},\chi_c)
\rangle$. The case $m=1$ reduces to the 2-point correlation function of the dark
matter overdensities along the \textit{los}, while the case $m=2$ 
requires the use of the 3-point correlation function or of its Fourier
counterpart, the bispectrum \cite{Bernardeau:2001qr}. In both cases we
also include in the analysis the fact that both the \Lya spectra and
the CMB convergence are being measured with a finite resolution. To
this end we add appropriate gaussian window functions that limit the
modes contributing to the CMB convergence (perpendicular to the los) to wavenumbers $|\vec{k}_{\perp}|\le k_C$ and
the ones contributing to the \Lya spectra (parallel to the los) to $k_l\le k_{\parallel}\le k_L$. When
these cutoffs are included, the treatment of the $m=2$ case becomes
involved. However, by expanding in power series the modified
Bessel functions arising from the angular integration in $k$--space, it
is possible to obtain an exact series solution for the
cumulant correlator.

We calculate the value of $\langle \delta\flux\,\kappa\rangle$ and of
$\langle \delta\flux^2\, \kappa\rangle$ as a function of the QSO
redshift $z$ and of the redshift range $\Delta z$ spanned by the \Lya
spectrum for a flat \lcdm cosmology with $\Omega_{\rm 0m}=0.25$,
$\sigma_8=0.84$, $n_{\rm s}=0.96$ and $H_0=72$ km/s/Mpc. The results
obtained -- shown in Fig.~\ref{Fig:S_3_2} -- make physical sense: if the \Lya spectrum covers a larger
redshift range the correlators increase, as more information is
carried by the longer spectrum. On the other hand, if the source QSO
redshift is increased while keeping the length of the spectrum fixed
the correlators become smaller because the \Lya spectrum is probing
regions where structure had less time to form. Furthermore, the rate
at which the correlators are increasing as a function of $\Delta z$ is
itself increasing with the CMB and \Lya experiment
resolutions $k_L$ and $k_C$. Finally, as expected the $\langle \delta
\flux^2\kappa\rangle$ correlator is about two orders of magnitude
larger than $\langle \delta\flux\, \kappa \rangle$: regions of higher
convergence have typically higher mean density and hence larger fluctuations' variance due to enhanced growth \cite{Zaldarriaga:2000}.

%AV: Second order perturbation theory
%implies that the variance is larger in dense
%regions. \textcolor{red}{(These last two sentences are cryptic. Can we
%  expand a bit on this?)}

%%%%%%%%%%%%%%%%%%%%%%%%%%%%%%%%%%%%%%%%%%%%%%%%
%%%
%%%%%%%%%%%%%%%
%%%%%%%%%%%%%%%%%%%%%%%%%%%%%%%%%%%%%%%%%%%%%%%%
%%%
%%%%%%%%%%%%%%%

\textit{Estimate for the correlators' signal-to-noise ratio.} To assess
the detectability of the above correlations by CMB
experiments and QSO surveys we need to evaluate their
signal--to--noise ratio (S/N). The calculation of the variances of
$\delta\flux\,\kappa$ and $\delta\flux^2\, \kappa$ requires the
evaluation of $\langle \delta\flux^2\,\kappa^2\rangle$ and of $\langle
\delta\flux^4\,\kappa^2\rangle$, which are proportional to $\langle
\delta_q\delta_{q'}\, \delta_c\delta_{c'} \rangle$ and $\langle
\delta_q^2\delta_{q'}^2 \delta_c\delta_{c'} \rangle$, respectively
[we adopt the notation
$\delta_i=\delta(\vec{x}_i)$]. The
exact evaluation of the latter requires the calculation of a six--point
function, the form of which to our knowledge has never been obtained. However, an \textit{estimate} for the dominant contributions
to the variance can be obtained using the approximation
%\begin{equation}
% \langle
%\delta_q^2\delta_{q'}^2 \delta_c\delta_{c'} %\rangle\approx2\langle\delta_q^2\delta_c\rangle^2
%+\langle\delta_q^2\delta_{q'}^2\rangle\langle \delta_c\delta_{c'} \rangle.
%\end{equation} 
\begin{eqnarray}
 \langle\delta^{2}_q\delta^{2}_{q'}\delta_c\delta_{c'}\rangle\approx
2\langle\delta^{2}_q\delta_c\rangle^2+\langle\delta_{c}\delta_{c'}\rangle\,\left(\langle\delta^2_{q}\rangle\langle\delta^2_{q'}\rangle+2\langle\delta_q\delta_{q'}\rangle^2\right),\label{N:d2qd2qdcdc}
\end{eqnarray}

\begin{figure*}[t!]
\includegraphics[width=0.45\textwidth]{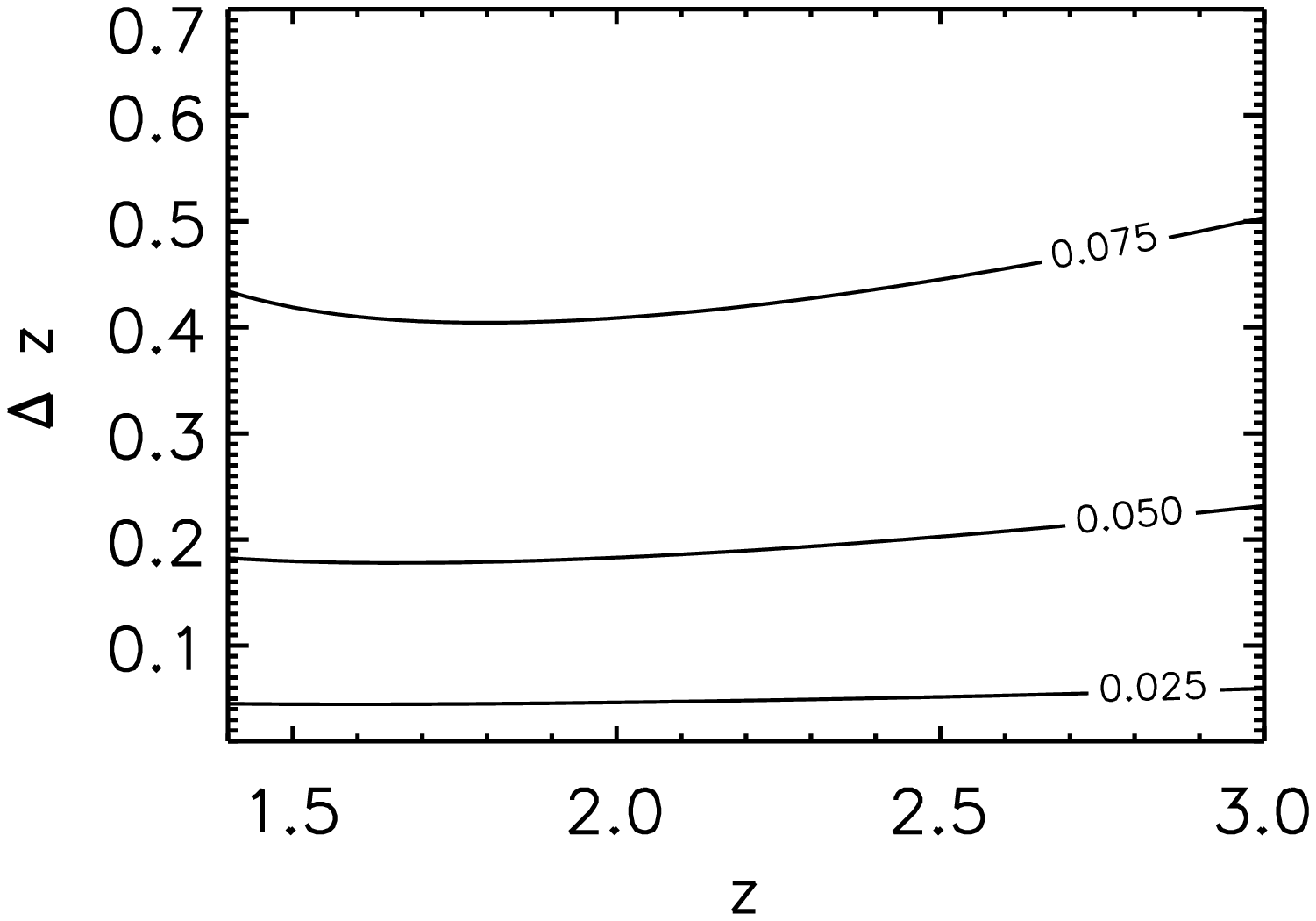}
\includegraphics[width=0.45\textwidth]{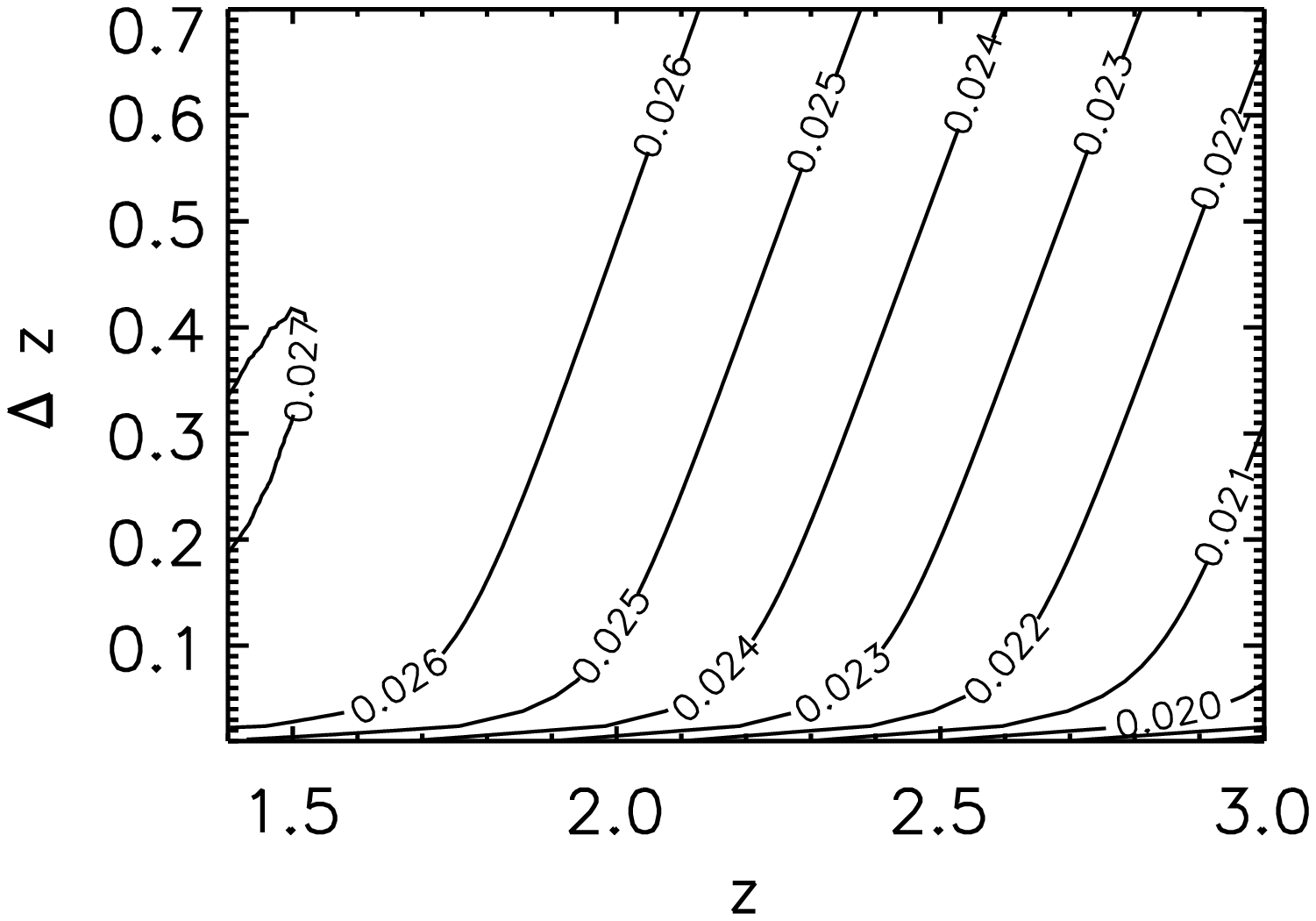}
\caption{Estimate of the S/N ratio for the
  measurement of the correlators $\langle \delta\flux \kappa\rangle$
  (left panel) and $\langle \delta \flux^2 \kappa\rangle$ (right
  panel) along a \textit{single} \textit{los} as a function of the source
  redshift $z$ and of the length of the measured spectrum $\Delta
  z$. The cutoff scales assumed are $k_l=0.01\hMpcI$ and
  $k_L=4.8\hMpcI$ for the QSO spectrum and $k_C=0.021\hMpcI$ for the
  CMB convergence, consistent with SDSS-III and Planck
  specifications.}
\label{Fig:SN_3_2}
\end{figure*}

The estimates for the S/N ratio for the measurement of
$\delta\flux\, \kappa$ and $\delta\flux^2\, \kappa$ along a
\textit{single} \textit{los} are shown in Fig.~\ref{Fig:SN_3_2}. As before,
increasing the length of the spectrum or decreasing the redshift of
the source will increase the S/N ratio, as more information is added
by the longer spectrum in the first case and as the shallower redshift
will allow to probe regions with clumpier structure in the second
case. Also, comparing the two panels it is possible to notice how the correlators'
different dependence on the growth of structure affects the redshift
range over which a given S/N ratio can be obtained. Finally, we also
notice that the S/N for $\delta\flux\, \kappa$ is about four times
larger than the one for $\delta\flux^2\, \kappa$, a consequence of the
fact that cosmic variance is generally higher for higher order
statistics.

%%%%%%%%%%%%%%%%%%%%%%%%%%%%%%%%%%%%%%%%%%%%%%%%
%%%
%%%%%%%%%%%%%%%
%%%%%%%%%%%%%%%%%%%%%%%%%%%%%%%%%%%%%%%%%%%%%%%%
%%%
%%%%%%%%%%%%%%%

%\begin{table}[b!]
%\begin{center}
%\begin{tabular}{ccc}
%\hline
%CMB Exp. & S/N per \textit{los} & Total S/N in BOSS\\
%\hline
%\hline
%Planck & 0.024 &  9.6\\
%PolExp & 0.05 & 20.0\\
%\hline
%\end{tabular}
%\end{center}
%\caption{Total and per single \textit{los} signal--to--noise (S/N) of the
%  $\delta\flux^2\,\kappa$ cross--correlation for
%  different CMB experiments combined with BOSS. For Planck, we adopt the sensitivity values of the 9 frequency channels from the Blue Book \cite{PlanckBB}. Here PolExp refers to
%  a hypothetical polarization based CMB experiment with a $3$ arcmin
%  beam and $800$ detectors, each having a noise-equivalent-temperature
%  (NET) of 300 $\mu$K-$\sqrt{s}$ over $8000$ sq. deg., with an
%  integration time of $3\times 10^7$ seconds.  The cross--correlation
%  is performed over the $8000$ sq. deg footprint of BOSS. The numbers
%  are calculated assuming that the mean QSO is situated at $z=2.5$
%  with $\Delta z =0.5$ and there are $20$ QSOs per square degree in
%  BOSS.}
%\label{table1}
%\end{table}%

\textit{Conclusions.} We presented a calculation of the
cross--correlation signal (and its noise) between the (variance of
the) \Lya flux fluctuations and the CMB convergence. We found that the
cross--correlation signal peaks at somewhat smaller redshifts than
those usually probed by \Lya alone, but it will be detectable with
future data sets with a large number of QSO. In particular, we estimate that the S/N ratio for $\langle\delta\flux^2\,\kappa\rangle$ obtained by cross-correlating the BOSS QSO sample with Planck data will yield a S/N ratio of 9.6. The use of higher definition data arising from a hypothetical CMB polarization experiment covering 8000 sq.~degrees would further increase the estimated S/N to 20.
%In Table \ref{table1} we
%estimate the S/N ratio for the $\delta\flux^2\,\kappa$ correlation for
%the BOSS QSO sample coupled with Planck or with a CMB polarization
%based experiment covering 8000 sq. degrees (PolExp). 
In all the above
cases the present analysis shows that this correlation 
%in the framework of a standard $\Lambda$CDM model 
will be observable. 
 
A few caveats are also in order. First, as large contributions to the
signal may actually be arising at small separations, it is necessary to stress that the \textit{estimates} obtained for the variance of the
correlators -- and consequently for S/N --
do not take into account the non--linearities
characteristic of small separations. The latter however are not
expected to alter significantly the results, given the QSO mean
redshift. 
Second, it is necessary to point out that the
\textit{estimates} of Fig.~\ref{Fig:SN_3_2} do not take into
account the detector systematics, which in a more realistic
calculation should also be included. 
Finally, we note that the
estimators considered may not be the ones
characterized by the least variance. It is reasonable to speculate
that the use of redshift weighted estimators in conjuction with lower
redshift convergence maps obtained through LSS surveys may allow an improvement in the S/N ratio. The formalism developed can further be extended to cross-correlate the \Lya flux with convergence maps obtained from optical lensing surveys once these will become available at the redshifts of the \Lya forest. This would result in an improvement in S/N ratio.

Finally, the present work is just a
first step in assessing the information content and the detectability
of \Lya--CMB lensing correlations and more refined calculations, based on
non--linear theory and hydrodynamical simulations should be performed
to address this more quantitatively, as future observational programs should enable to perform such a measurement with several consequences. First, this would provide independent constraints on the bias between flux and matter, additional to the ones that are obtained by high resolution QSO spectra and hydrodynamical simulations. Second, being sensitive to the growth of structures at intermediate to small scales, this correlation could be used to test for models of early dark energy \cite{Xia:2009ys} or modification of gravity that produce scale dependent growth evolutions \cite{Zhao:2008bn} and to provide constraints on the neutrino masses.

%new astrophysical insights on the flux--matter
%relation and cosmological constraints on the growth of structures
%in the high--redshift universe.
% in either standard or modified models of structure formation.

%{AV: Here we need to answer the questions: what do we learn from measuring $\delta\flux^2 \kappa$? What is it that we better understand/constrain? Let's give some perspective.}

\textit{Acknowledgements:} We thank F.~Bernardeau, S.~Matarrese, S.~Dodelson,
J.~Frieman, E.~Sefusatti, N.~Gnedin and J.~P.~Uzan for useful
conversations. AV is supported by the DOE and the NASA grant NAG
5-10842 at Fermilab.  SD thanks APC - Paris for hospitality when this
project was conceived. SD and DS acknowledge NASA grant NNX08AH30G and
NSF grant 0707731. MV acknowledges support from an ASI/AAE grant.

\bibliography{VDSV,cmblensing,projects_new}

\begin{thebibliography}{31}
\expandafter\ifx\csname natexlab\endcsname\relax\def\natexlab#1{#1}\fi
\expandafter\ifx\csname bibnamefont\endcsname\relax
  \def\bibnamefont#1{#1}\fi
\expandafter\ifx\csname bibfnamefont\endcsname\relax
  \def\bibfnamefont#1{#1}\fi
\expandafter\ifx\csname citenamefont\endcsname\relax
  \def\citenamefont#1{#1}\fi
\expandafter\ifx\csname url\endcsname\relax
  \def\url#1{\texttt{#1}}\fi
\expandafter\ifx\csname urlprefix\endcsname\relax\def\urlprefix{URL }\fi
\providecommand{\bibinfo}[2]{#2}
\providecommand{\eprint}[2][]{\url{#2}}

\bibitem[{\citenamefont{{McDonald}
  et~al.}(2005{\natexlab{a}})\citenamefont{{McDonald}, {Seljak}, {Cen}, {Shih},
  {Weinberg}, {Burles}, {Schneider}, {Schlegel}, {Bahcall}, {Briggs}
  et~al.}}]{mcdonald05}
\bibinfo{author}{\bibfnamefont{P.}~\bibnamefont{{McDonald}}},
  \bibinfo{author}{\bibfnamefont{U.}~\bibnamefont{{Seljak}}},
  \bibinfo{author}{\bibfnamefont{R.}~\bibnamefont{{Cen}}},
  \bibinfo{author}{\bibfnamefont{D.}~\bibnamefont{{Shih}}},
  \bibinfo{author}{\bibfnamefont{D.~H.} \bibnamefont{{Weinberg}}},
  \bibinfo{author}{\bibfnamefont{S.}~\bibnamefont{{Burles}}},
  \bibinfo{author}{\bibfnamefont{D.~P.} \bibnamefont{{Schneider}}},
  \bibinfo{author}{\bibfnamefont{D.~J.} \bibnamefont{{Schlegel}}},
  \bibinfo{author}{\bibfnamefont{N.~A.} \bibnamefont{{Bahcall}}},
  \bibinfo{author}{\bibfnamefont{J.~W.} \bibnamefont{{Briggs}}},
  \bibnamefont{et~al.}, \bibinfo{journal}{\apj} \textbf{\bibinfo{volume}{635}},
  \bibinfo{pages}{761} (\bibinfo{year}{2005}{\natexlab{a}}),
  \eprint{arXiv:astro-ph/0407377}.
\bibinfo{author}{\bibfnamefont{U.}~\bibnamefont{{Seljak}}},
  \bibinfo{author}{\bibfnamefont{A.}~\bibnamefont{{Makarov}}},
  \bibinfo{author}{\bibfnamefont{P.}~\bibnamefont{{McDonald}}},
  \bibinfo{author}{\bibfnamefont{S.~F.} \bibnamefont{{Anderson}}},
  \bibinfo{author}{\bibfnamefont{N.~A.} \bibnamefont{{Bahcall}}},
  \bibinfo{author}{\bibfnamefont{J.}~\bibnamefont{{Brinkmann}}},
  \bibinfo{author}{\bibfnamefont{S.}~\bibnamefont{{Burles}}},
  \bibinfo{author}{\bibfnamefont{R.}~\bibnamefont{{Cen}}},
  \bibinfo{author}{\bibfnamefont{M.}~\bibnamefont{{Doi}}},
  \bibinfo{author}{\bibfnamefont{J.~E.} \bibnamefont{{Gunn}}},
  \bibnamefont{et~al.}, \bibinfo{journal}{\prd} \textbf{\bibinfo{volume}{71}},
  \bibinfo{pages}{103515} (\bibinfo{year}{2005}),
  \eprint{arXiv:astro-ph/0407372}.
\bibinfo{author}{\bibfnamefont{D.}~\bibnamefont{Schlegel}},
  \bibinfo{author}{\bibfnamefont{M.}~\bibnamefont{White}}, \bibnamefont{and}
  \bibinfo{author}{\bibfnamefont{D.}~\bibnamefont{Eisenstein}}
  (\bibinfo{collaboration}{with input from the SDSS-III})
  (\bibinfo{year}{2009}), \eprint{0902.4680}.

\bibitem[{\citenamefont{{Gratton} et~al.}(2008)\citenamefont{{Gratton},
  {Lewis}, and {Efstathiou}}}]{gratton.lewis.ea:2008}
\bibinfo{author}{\bibfnamefont{S.}~\bibnamefont{{Gratton}}},
  \bibinfo{author}{\bibfnamefont{A.}~\bibnamefont{{Lewis}}}, \bibnamefont{and}
  \bibinfo{author}{\bibfnamefont{G.}~\bibnamefont{{Efstathiou}}},
  \bibinfo{journal}{\prd} \textbf{\bibinfo{volume}{77}},
  \bibinfo{pages}{083507} (\bibinfo{year}{2008}), \eprint{0705.3100}.

\bibitem[{\citenamefont{{Lesgourgues} et~al.}(2007)\citenamefont{{Lesgourgues},
  {Viel}, {Haehnelt}, and {Massey}}}]{lesgourgues07}
\bibinfo{author}{\bibfnamefont{J.}~\bibnamefont{{Lesgourgues}}},
  \bibinfo{author}{\bibfnamefont{M.}~\bibnamefont{{Viel}}},
  \bibinfo{author}{\bibfnamefont{M.~G.} \bibnamefont{{Haehnelt}}},
  \bibnamefont{and} \bibinfo{author}{\bibfnamefont{R.}~\bibnamefont{{Massey}}},
  \bibinfo{journal}{Journal of Cosmology and Astro-Particle Physics}
  \textbf{\bibinfo{volume}{11}}, \bibinfo{pages}{8} (\bibinfo{year}{2007}),
  \eprint{0705.0533}.

\bibitem[{\citenamefont{{Seljak} and {Slosar}}(2006)}]{seljak.slosar:2006}
\bibinfo{author}{\bibfnamefont{U.}~\bibnamefont{{Seljak}}} \bibnamefont{and}
  \bibinfo{author}{\bibfnamefont{A.}~\bibnamefont{{Slosar}}},
  \bibinfo{journal}{\prd} \textbf{\bibinfo{volume}{74}},
  \bibinfo{pages}{063523} (\bibinfo{year}{2006}),
  \eprint{arXiv:astro-ph/0604143}.

\bibitem[{\citenamefont{{McDonald} and
  {Eisenstein}}(2007)}]{mcdonald.eisentein:2007}
\bibinfo{author}{\bibfnamefont{P.}~\bibnamefont{{McDonald}}} \bibnamefont{and}
  \bibinfo{author}{\bibfnamefont{D.~J.} \bibnamefont{{Eisenstein}}},
  \bibinfo{journal}{\prd} \textbf{\bibinfo{volume}{76}},
  \bibinfo{pages}{063009} (\bibinfo{year}{2007}),
  \eprint{arXiv:astro-ph/0607122}.

\bibitem[{\citenamefont{{Viel} et~al.}(2008)\citenamefont{{Viel}, {Becker},
  {Bolton}, {Haehnelt}, {Rauch}, and {Sargent}}}]{viel08}
\bibinfo{author}{\bibfnamefont{M.}~\bibnamefont{{Viel}}},
  \bibinfo{author}{\bibfnamefont{G.~D.} \bibnamefont{{Becker}}},
  \bibinfo{author}{\bibfnamefont{J.~S.} \bibnamefont{{Bolton}}},
  \bibinfo{author}{\bibfnamefont{M.~G.} \bibnamefont{{Haehnelt}}},
  \bibinfo{author}{\bibfnamefont{M.}~\bibnamefont{{Rauch}}}, \bibnamefont{and}
  \bibinfo{author}{\bibfnamefont{W.~L.~W.} \bibnamefont{{Sargent}}},
  \bibinfo{journal}{Physical Review Letters} \textbf{\bibinfo{volume}{100}},
  \bibinfo{pages}{041304} (\bibinfo{year}{2008}), \eprint{0709.0131}.

\bibitem[{\citenamefont{Viel et~al.}(2004)\citenamefont{Viel, Haehnelt, and
  Springel}}]{vhs04}
\bibinfo{author}{\bibfnamefont{M.}~\bibnamefont{Viel}},
  \bibinfo{author}{\bibfnamefont{M.~G.} \bibnamefont{Haehnelt}},
  \bibnamefont{and} \bibinfo{author}{\bibfnamefont{V.}~\bibnamefont{Springel}},
  \bibinfo{journal}{Mon. Not. Roy. Astron. Soc.}
  \textbf{\bibinfo{volume}{354}}, \bibinfo{pages}{684} (\bibinfo{year}{2004}),
  \eprint{astro-ph/0404600}.
\bibinfo{author}{\bibfnamefont{P.}~\bibnamefont{{McDonald}}},
  \bibinfo{author}{\bibfnamefont{U.}~\bibnamefont{{Seljak}}},
  \bibinfo{author}{\bibfnamefont{R.}~\bibnamefont{{Cen}}},
  \bibinfo{author}{\bibfnamefont{P.}~\bibnamefont{{Bode}}}, \bibnamefont{and}
  \bibinfo{author}{\bibfnamefont{J.~P.} \bibnamefont{{Ostriker}}},
  \bibinfo{journal}{\mnras} \textbf{\bibinfo{volume}{360}},
  \bibinfo{pages}{1471} (\bibinfo{year}{2005}{\natexlab{b}}),
  \eprint{arXiv:astro-ph/0407378}.
\bibinfo{author}{\bibfnamefont{J.~S.} \bibnamefont{{Bolton}}},
  \bibinfo{author}{\bibfnamefont{M.}~\bibnamefont{{Viel}}},
  \bibinfo{author}{\bibfnamefont{T.-S.} \bibnamefont{{Kim}}},
  \bibinfo{author}{\bibfnamefont{M.~G.} \bibnamefont{{Haehnelt}}},
  \bibnamefont{and} \bibinfo{author}{\bibfnamefont{R.~F.}
  \bibnamefont{{Carswell}}}, \bibinfo{journal}{\mnras}
  \textbf{\bibinfo{volume}{386}}, \bibinfo{pages}{1131} (\bibinfo{year}{2008}),
  \eprint{0711.2064}.

\bibitem[{\citenamefont{{Lewis} and {Challinor}}(2006)}]{lewis.challinor:2006}
\bibinfo{author}{\bibfnamefont{A.}~\bibnamefont{{Lewis}}} \bibnamefont{and}
  \bibinfo{author}{\bibfnamefont{A.}~\bibnamefont{{Challinor}}},
  \bibinfo{journal}{\physrep} \textbf{\bibinfo{volume}{429}},
  \bibinfo{pages}{1} (\bibinfo{year}{2006}), \eprint{arXiv:astro-ph/0601594}.

\bibitem[{\citenamefont{{Hu} and {Okamoto}}(2002)}]{hu.okamoto:2002}
\bibinfo{author}{\bibfnamefont{W.}~\bibnamefont{{Hu}}} \bibnamefont{and}
  \bibinfo{author}{\bibfnamefont{T.}~\bibnamefont{{Okamoto}}},
  \bibinfo{journal}{\apj} \textbf{\bibinfo{volume}{574}}, \bibinfo{pages}{566}
  (\bibinfo{year}{2002}), \eprint{arXiv:astro-ph/0111606}.
\bibinfo{author}{\bibfnamefont{C.~M.} \bibnamefont{{Hirata}}} \bibnamefont{and}
  \bibinfo{author}{\bibfnamefont{U.}~\bibnamefont{{Seljak}}},
  \bibinfo{journal}{\prd} \textbf{\bibinfo{volume}{68}},
  \bibinfo{pages}{083002} (\bibinfo{year}{2003}),
  \eprint{arXiv:astro-ph/0306354}.
\bibinfo{author}{\bibfnamefont{J.}~\bibnamefont{{Yoo}}} \bibnamefont{and}
  \bibinfo{author}{\bibfnamefont{M.}~\bibnamefont{{Zaldarriaga}}},
  \bibinfo{journal}{ArXiv e-prints} \eprint{0805.2155}.

\bibitem[{\citenamefont{{Smith} et~al.}(2007)\citenamefont{{Smith}, {Zahn}, and
  {Dore}}}]{smith.zahn.ea:2007}
\bibinfo{author}{\bibfnamefont{K.~M.} \bibnamefont{{Smith}}},
  \bibinfo{author}{\bibfnamefont{O.}~\bibnamefont{{Zahn}}}, \bibnamefont{and}
  \bibinfo{author}{\bibfnamefont{O.}~\bibnamefont{{Dore}}},
  \bibinfo{journal}{\prd} \textbf{\bibinfo{volume}{76}},
  \bibinfo{pages}{043510}
  (\bibinfo{year}{2007}), \eprint{0705.3980}.

\bibitem[{\citenamefont{{Hirata}
  et~al.}(2008{\natexlab{b}})\citenamefont{{Hirata}, {Ho}, {Padmanabhan},
  {Seljak}, and {Bahcall}}}]{hirata08}
\bibinfo{author}{\bibfnamefont{C.~M.} \bibnamefont{{Hirata}}},
  \bibinfo{author}{\bibfnamefont{S.}~\bibnamefont{{Ho}}},
  \bibinfo{author}{\bibfnamefont{N.}~\bibnamefont{{Padmanabhan}}},
  \bibinfo{author}{\bibfnamefont{U.}~\bibnamefont{{Seljak}}}, \bibnamefont{and}
  \bibinfo{author}{\bibfnamefont{N.~A.} \bibnamefont{{Bahcall}}},
  \bibinfo{journal}{\prd} \textbf{\bibinfo{volume}{78}},
  \bibinfo{pages}{043520} (\bibinfo{year}{2008}{\natexlab{b}}),
  \eprint{0801.0644}.

\bibitem[{ALL()}]{ALL}
\bibinfo{note}{\url{http://www.physics.princeton.edu/act};}
\bibinfo{note}{\url{http://quiet.uchicago.edu/};}
\bibinfo{note}{\url{http://pole.uchicago.edu};\url{http://www.rssd.esa.int/index.php?project=planck};}
\bibinfo{note}{\url{http://www.sdss3.org/cosmology.php};\url{http://bolo.berkeley.edu/polarbear/}}.

\bibitem[{\citenamefont{{Peiris} and {Spergel}}(2000)}]{peiris.spergel:2000}
\bibinfo{author}{\bibfnamefont{H.~V.} \bibnamefont{{Peiris}}} \bibnamefont{and}
  \bibinfo{author}{\bibfnamefont{D.~N.} \bibnamefont{{Spergel}}},
  \bibinfo{journal}{\apj} \textbf{\bibinfo{volume}{540}}, \bibinfo{pages}{605}
  (\bibinfo{year}{2000}), \eprint{arXiv:astro-ph/0001393}.
\bibinfo{author}{\bibfnamefont{T.}~\bibnamefont{{Giannantonio}}},
  \bibinfo{author}{\bibfnamefont{R.}~\bibnamefont{{Scranton}}},
  \bibinfo{author}{\bibfnamefont{R.~G.} \bibnamefont{{Crittenden}}},
  \bibinfo{author}{\bibfnamefont{R.~C.} \bibnamefont{{Nichol}}},
  \bibinfo{author}{\bibfnamefont{S.~P.} \bibnamefont{{Boughn}}},
  \bibinfo{author}{\bibfnamefont{A.~D.} \bibnamefont{{Myers}}},
  \bibnamefont{and} \bibinfo{author}{\bibfnamefont{G.~T.}
  \bibnamefont{{Richards}}}, \bibinfo{journal}{\prd}
  \textbf{\bibinfo{volume}{77}}, \bibinfo{pages}{123520}
  (\bibinfo{year}{2008}), \eprint{0801.4380}.
\bibinfo{author}{\bibfnamefont{S.}~\bibnamefont{{Ho}}},
  \bibinfo{author}{\bibfnamefont{C.}~\bibnamefont{{Hirata}}},
  \bibinfo{author}{\bibfnamefont{N.}~\bibnamefont{{Padmanabhan}}},
  \bibinfo{author}{\bibfnamefont{U.}~\bibnamefont{{Seljak}}}, \bibnamefont{and}
  \bibinfo{author}{\bibfnamefont{N.}~\bibnamefont{{Bahcall}}},
  \bibinfo{journal}{\prd} \textbf{\bibinfo{volume}{78}},
  \bibinfo{pages}{043519} (\bibinfo{year}{2008}), \eprint{0801.0642}.
\bibinfo{author}{\bibfnamefont{R.~A.~C.}~\bibnamefont{{Croft}}},
  \bibinfo{author}{\bibfnamefont{A.~J.}~\bibnamefont{{Banday}}}, \bibnamefont{and}
  \bibinfo{author}{\bibfnamefont{L.}~\bibnamefont{{Hernquist}}},
  \bibinfo{journal}{Mon. Not. Roy. Astron. Soc.} \textbf{\bibinfo{volume}{369}},
  \bibinfo{pages}{1090} (\bibinfo{year}{2006}), \eprint{astro-ph/0512380}.

\bibitem[{\citenamefont{Zaldarriaga et~al.}(2000)\citenamefont{Zaldarriaga, Seljak, and Hui}}]{Zaldarriaga:2000}
\bibinfo{author}{\bibfnamefont{M.}~\bibnamefont{Zaldarriaga}},
  \bibinfo{author}{\bibfnamefont{U.}~\bibnamefont{Seljak}},
  \bibnamefont{and}
  \bibinfo{author}{\bibfnamefont{L.}~\bibnamefont{Hui}},
  \bibinfo{journal}{Astrophys. J.} \textbf{\bibinfo{volume}{551}},
  \bibinfo{pages}{48} (\bibinfo{year}{2001}), \eprint{astro-ph/0007101}.

\bibitem[{\citenamefont{Hui and Gnedin}(1997)}]{huignedin97}
\bibinfo{author}{\bibfnamefont{L.}~\bibnamefont{Hui}} \bibnamefont{and}
  \bibinfo{author}{\bibfnamefont{N.~Y.} \bibnamefont{Gnedin}},
  \bibinfo{journal}{Mon. Not. Roy. Astron. Soc.}
  \textbf{\bibinfo{volume}{292}}, \bibinfo{pages}{27} (\bibinfo{year}{1997}),
  \eprint{astro-ph/9612232}.

\bibitem[{\citenamefont{Gunn and Peterson}(1965)}]{Gunn:1965hd}
\bibinfo{author}{\bibfnamefont{J.~E.} \bibnamefont{Gunn}} \bibnamefont{and}
  \bibinfo{author}{\bibfnamefont{B.~A.} \bibnamefont{Peterson}},
  \bibinfo{journal}{Astrophys. J.} \textbf{\bibinfo{volume}{142}},
  \bibinfo{pages}{1633} (\bibinfo{year}{1965}).
\bibinfo{author}{\bibfnamefont{H.}~\bibnamefont{Bi}} \bibnamefont{and}
  \bibinfo{author}{\bibfnamefont{A.~F.} \bibnamefont{Davidsen}},
  \bibinfo{journal}{Astrophys. J.} \textbf{\bibinfo{volume}{479}},
  \bibinfo{pages}{523} (\bibinfo{year}{1997}), \eprint{astro-ph/9611062}.

\bibitem[{\citenamefont{Croft et~al.}(1998)\citenamefont{Croft, Weinberg, Katz,
  and Hernquist}}]{Croft:1997jf}
\bibinfo{author}{\bibfnamefont{R.~A.~C.} \bibnamefont{Croft}},
  \bibinfo{author}{\bibfnamefont{D.~H.} \bibnamefont{Weinberg}},
  \bibinfo{author}{\bibfnamefont{N.}~\bibnamefont{Katz}}, \bibnamefont{and}
  \bibinfo{author}{\bibfnamefont{L.}~\bibnamefont{Hernquist}},
  \bibinfo{journal}{Astrophys. J.} \textbf{\bibinfo{volume}{495}},
  \bibinfo{pages}{44} (\bibinfo{year}{1998}), \eprint{astro-ph/9708018}.

\bibitem[{\citenamefont{{McDonald}}(2003)}]{mcdonald03}
\bibinfo{author}{\bibfnamefont{P.}~\bibnamefont{{McDonald}}},
  \bibinfo{journal}{\apj} \textbf{\bibinfo{volume}{585}}, \bibinfo{pages}{34}
  (\bibinfo{year}{2003}), \eprint{arXiv:astro-ph/0108064}.

\bibitem[{\citenamefont{Viel et~al.}(2002)\citenamefont{Viel, Matarrese, Mo,
  Haehnelt, and Theuns}}]{Viel:2001hd}
\bibinfo{author}{\bibfnamefont{M.}~\bibnamefont{Viel}},
  \bibinfo{author}{\bibfnamefont{S.}~\bibnamefont{Matarrese}},
  \bibinfo{author}{\bibfnamefont{H.~J.} \bibnamefont{Mo}},
  \bibinfo{author}{\bibfnamefont{M.~G.} \bibnamefont{Haehnelt}},
  \bibnamefont{and} \bibinfo{author}{\bibfnamefont{T.}~\bibnamefont{Theuns}},
  \bibinfo{journal}{Mon. Not. Roy. Astron. Soc.}
  \textbf{\bibinfo{volume}{329}}, \bibinfo{pages}{848} (\bibinfo{year}{2002}),
  \eprint{astro-ph/0105233}.

\bibitem[{\citenamefont{Bernardeau et~al.}(2002)\citenamefont{Bernardeau,
  Colombi, Gaztanaga, and Scoccimarro}}]{Bernardeau:2001qr}
\bibinfo{author}{\bibfnamefont{F.}~\bibnamefont{Bernardeau}},
  \bibinfo{author}{\bibfnamefont{S.}~\bibnamefont{Colombi}},
  \bibinfo{author}{\bibfnamefont{E.}~\bibnamefont{Gaztanaga}},
  \bibnamefont{and}
  \bibinfo{author}{\bibfnamefont{R.}~\bibnamefont{Scoccimarro}},
  \bibinfo{journal}{Phys. Rept.} \textbf{\bibinfo{volume}{367}},
  \bibinfo{pages}{1} (\bibinfo{year}{2002}), \eprint{astro-ph/0112551}.


\bibitem[{\citenamefont{Xia and Viel}(2009)}]{Xia:2009ys}
\bibinfo{author}{\bibfnamefont{J.-Q.} \bibnamefont{Xia}} \bibnamefont{and}
  \bibinfo{author}{\bibfnamefont{M.}~\bibnamefont{Viel}},
  \bibinfo{journal}{JCAP} \textbf{\bibinfo{volume}{0904}}, \bibinfo{pages}{002}
  (\bibinfo{year}{2009}), \eprint{0901.0605}.

\bibitem[{\citenamefont{Zhao et~al.}(2008)\citenamefont{Zhao, Pogosian,
  Silvestri, and Zylberberg}}]{Zhao:2008bn}
\bibinfo{author}{\bibfnamefont{G.-B.} \bibnamefont{Zhao}},
  \bibinfo{author}{\bibfnamefont{L.}~\bibnamefont{Pogosian}},
  \bibinfo{author}{\bibfnamefont{A.}~\bibnamefont{Silvestri}},
  \bibnamefont{and}
  \bibinfo{author}{\bibfnamefont{J.}~\bibnamefont{Zylberberg}}
  (\bibinfo{year}{2008}), \eprint{0809.3791}.

\end{thebibliography}

\end{document}